\documentclass{jfm}
\usepackage{graphicx}
\usepackage{epstopdf, epsfig}
\usepackage{bm}
\usepackage{natbib}
\usepackage{ar}
\usepackage{amsmath}
\usepackage{amssymb}
\usepackage{rotating}
\usepackage{pdflscape}
\usepackage{afterpage}

\newcommand{\vect}[1]{\textbf{\textit{#1}}} 
\newcommand{\vort}{\boldsymbol{\omega}} 
\newcommand{\uv}[1]{\hat{\boldsymbol{#1}}} 
\newcommand{\der}[2]{\frac{\mathrm{d}#1}{\mathrm{d}#2}} 
\newcommand{\pder}[2]{\frac{\partial#1}{\partial#2}} 
\newcommand{\del}{\boldsymbol{\nabla}} 
\newcommand{\grad}[1]{\del #1} 
\newcommand{\mdiv}[1]{\del\cdot\boldsymbol{#1}} 
\newcommand{\curl}[1]{\del\times\boldsymbol{#1}} 
\newcommand{\md}[1]{\mathrm{d}#1} 
\newcommand{\mdi}[1]{\hspace{2pt}\mathrm{d}#1} 
\newcommand{\jb}[1]{[\![#1]\!]} 
\newcommand{\e}[1]{\mathrm{e}^{#1}} 
\newcommand{\gbf}[1]{\boldsymbol{#1}} 

\newcommand{\lrp}[1]{\left(#1\right)} 
\newcommand{\bp}[1]{\big(#1\big)} 
\newcommand{\lrb}[1]{\left[#1\right]} 
\newcommand{\Bb}[1]{\Big[#1\Big]} 

\shorttitle{The force on a body moving in an inviscid fluid}
\shortauthor{A. C. DeVoria and K. Mohseni}

\title{The force on a body moving in an inviscid fluid}
\author{A. C. DeVoria\aff{1} \and K. Mohseni\aff{1,2}\corresp{\email{mohseni@ufl.edu}}}

\affiliation{\aff{1}Department of Mechanical \& Aerospace Engineering, University of Florida, Gainesville, FL 32611, USA
\aff{2}Department of Electrical \& Computer Engineering, University of Florida, Gainesville, FL 32611, USA}

\begin{document}

\maketitle

\begin{abstract}
This paper presents some novel contributions to the theory of inviscid flow regarding the forces exerted on a body moving through such a fluid in two dimensions. It is argued that acceleration of the body corresponds to vorticity generation that is independent of the instantaneous velocity of the body and thus the boundary condition on the normal velocity. The strength of the vortex sheet representing the body retains a degree of freedom that represents the net effect of the tangential boundary condition associated with the viscous flow governed by the higher-order Navier-Stokes equations. This degree of freedom is the circulation of the vorticity generated by the acceleration of the body. Equivalently, it is the net circulation around a contour enclosing the body and any shed vorticity. In accordance with Kelvin's circulation theorem, a non-zero value of the circulation around this system is necessarily communicated to infinity. This contrasts with the usual acceptance of the theorem as requiring this circulation to be zero at all times; a condition that is incapable of capturing the effect of newly generated vorticity on the body surface when it accelerates. Additionally, the usual boundary condition of continuity of normal velocity is relaxed to allow for fluid entrainment into surfaces of discontinuity that represent the mass contained within the viscous layers of the physical problem. The generalized force calculation is presented in detail. The importance of the vorticity generation due to body acceleration is demonstrated on some modeled problems relevant to biological propulsion. For fast, fish-like locomotion with aggressive fin movements, asymmetry in the form of a propulsive stroke results in a net thrust. For smooth undulatory locomotion, thrust is generated by a phase shift between the circulation production and the fin motion. The modeled thrust generation is also compared to experimental measurements on oscillating foils and shows fair agreement. 
\end{abstract}

\section{Introduction}\label{sec:intro}
Potential and inviscid flow models have been and continue to be profitable tools for prediction in fluid mechanics, even as increasing computing power readily puts numerical solutions of the full Navier-Stokes equations at our fingertips. Despite some mathematical restrictions of the equations governing the simplified fluids, researchers continue to successfully employ such methods to new applications as well as to improve upon modeling of longstanding problems. 

A model of the flow field for a given problem is of obvious interest to the academic, while the application-driven or practical interest is often focused on obtaining engineering quantities, such as the force on a body moving within the fluid. However, in inviscid flow modeling little has changed in regard to this force calculation. One reason for this stems from a key assumption that is imposed on the problem, often as an axiom, from the onset. The assumption concerns the application of Kelvin's circulation theorem for an inviscid fluid and specifically its effect on vorticity generation at the surface of an accelerating body. The purpose of this introduction is to further elucidate this issue. 

Before continuing in this vein, however, we briefly draw attention to a recent work by \cite{GrahamWR:19a}, in which the author provides an excellent expos{\'e} of the force decomposition for a body moving in a viscous, incompressible fluid. In particular, the pressure field is decomposed into viscous and inviscid parts. The latter gives rise to separate `convective' and `accelerative' forces that are more intuitive alternatives to the `added-mass' forces associated with the conventional decomposition of the velocity field into irrotational and rotational parts. In this paper, we will encounter similar concepts as those endorsed by Graham and we encourage the interested reader to explore his paper as well.

We begin rather generally with the consideration of a body moving through a viscous compressible fluid of unbounded extent in three-dimensional space and which is at rest at infinity. By a general theorem in classical hydrodynamics, the flow field is \textit{uniquely} determined by the vorticity distribution, $\vort=\curl{u}$, the expansion or dilatation distribution, $\Delta=\mdiv{u}$, and specification of the normal component of velocity on the boundary \citep{Lamb:45a}. If the region is multiply-connected, then the circulation (cyclic constant) around any irreducible curves must also be given. The additional enforcement of a boundary condition on the tangential components of velocity determines the strength of a vortex sheet on the body surface, which then immediately enters the fluid domain as a non-singular distribution by the process of diffusion, with the sequence of these events occurring in the order stated \citep{LighthillMJ:63a}. Accordingly, there is a natural `bookkeeping' of the vorticity entering the domain. In particular, the no-slip condition restricts the class of vorticity distributions that correspond to physically `relevant' flows. While there is nearly irrefutable evidence of the correctness of the no-slip condition under ordinary circumstances, it must be borne in mind that it is not a fundamental principle, but a phenomenological one that is used to `select' a unique flow solution (i.e. vorticity distribution). Stated differently, the higher-order equation governing a viscous fluid is able to accommodate a boundary condition on both the normal and tangential velocity components, with the latter replacing the requirement of \textit{a priori} knowledge of the vorticity distribution (aside from an initial condition), which in the absence of non-conservative forces (e.g. baroclinic torque) must originate from the boundaries. For cases with very high rates of shear and strain, actual molecular slip may occur at the boundary, and thus the flow evolves according to a more generalized tangential velocity boundary condition \citep[e.g. see][]{Mohseni:16b}.

Now consider the inviscid version of the problem described above. The lower-order equation governing this fluid cannot satisfy a desired condition on the tangential velocity at all points of the boundary. Since diffusion has been eliminated as a transport mechanism, we can view the differential equation as having been integrated to `collapse' the vorticity to the boundary where it remains as a vortex sheet, and `shedding' of the sheet from the body is really the extension of the boundary into the fluid, albeit with this portion having its own different dynamics. As such, the tangential boundary condition is also `collapsed' from a (possibly varying) function along the boundary to a scalar degree of freedom. In exactly the same way that the no-slip condition selects a particular, unique local vorticity generation, the specification of this scalar selects a particular, unique local vortex sheet strength. Similarly, there is no fundamental principle to objectively decide the correct value corresponding to the relevant flow. The best we can hope for is a condition that captures the net or integrated effect of the local vorticity generation at the boundary as determined by the tangential boundary condition. That we must provide this condition reflects the fact that we have also eliminated the bookkeeping of vorticity entering the fluid domain via diffusion.

For familiarity we move forward with an incompressible fluid with $\Delta\equiv 0$ everywhere except possibly on sheets of discontinuity or other irregular points. Also, we shall constrain ourselves to a two-dimensional problem so that the presence of the body makes the region doubly-connected. As such, the scalar degree of freedom remaining in the local tangential boundary condition mentioned above can be taken as the global circulation around the body. Our purpose here is to the detail the way in which the effect of viscous stress on the body surface can be taken into account by an inviscid flow model whereby a vortex sheet is generated on, but does not diffuse from the body. We will assume that the fluid-surface interaction model satisfies the no-slip condition as well as continuity of stress; more generalized surface models can be found in \cite{Mohseni:19d} and \cite{Mohseni:20b}. Under these assumptions, the total stress at the surface of the body is equal to that of the fluid and is:
\begin{eqnarray}\label{eqn:stress}
\gbf{\sigma} = -p\uv{n}-\mu\uv{n}\times\vort = -p\uv{n}-\mu\omega\uv{s}
\end{eqnarray}
where $p$ is the pressure, $\mu$ is the dynamic viscosity of the fluid, and $\omega$ is the out-of-plane component of vorticity. The result for three-dimensional flow of a compressible fluid can be found in \cite{WuJZ:98a}. Integrating the stress over the body contour $C_b$ yields the total exerted force:
\begin{eqnarray}\label{eqn:force}
\vect{F}&=&-\oint_{C_b}p\uv{n}\mdi{l}-\mu\oint_{C_b}\omega\uv{s}\mdi{l}=\vect{F}_p+\vect{F}_\text{vis}.
\end{eqnarray}
The viscous force, by definition, is related to the vorticity on the surface. If we are in complete ignorance of the viscous effects and take $\vect{F}_\text{vis}=\textbf{0}$ by $\mu=0$, then not only will no vorticity enter the fluid, it will never even be created; the consequences of this are well known from classical potential theory. If we are to represent such viscous effects by the generation of a vortex sheet at the surface, then the tangential boundary condition, i.e. no-slip, must be imposed in some manner so that the net transfer of tangential momentum at the boundary is preserved. This subtle difference is that between an Euler solution and an Euler-limit of a Navier-Stokes solution \citep{LagerstromPA:75a,WuJZ:93a}. In other words, we are tasked with finding the relevant Euler solution with surfaces of discontinuity, i.e. vortex sheets, such that the contribution to the pressure $p$ from the induced velocity of the sheets retains the effect of $\vect{F}_\text{vis}$ upon setting $\mu$ to zero. 

So far, we have not really said anything new about the process of determining the vortex sheet strength at the surface, which is accomplished, in part, by satisfying the normal boundary condition on the body. To this end, consider the sheet strength $\gamma_b$ that has been generated by some movement of the body and any other induced flow at its surface. We let viscosity act on the sheet for a time $\Delta t\ll 1$ so that the vorticity is regularized into a thin layer around the body as \citep{LighthillMJ:63a}: 
\begin{equation}\label{eqn:vort}
\omega(s,n,\Delta t) = \frac{\gamma_b(s)}{\sqrt{4\pi\nu\Delta t}}\hspace{4pt}\mathrm{exp}\left\{\frac{-n^2}{4\nu\Delta t}\right\}
\end{equation}
where the body-tangent and body-normal coordinates are $s$ and $n$. Integration of $\omega$ from the body to a distance well outside the layer (i.e. infinity) gives $\gamma_b(s)$ as the conserved vorticity `source' strength. Substituting the above into (\ref{eqn:force}) yields an estimate for the net viscous force on the body ($n=0$) due to shear stress as:
\begin{equation}\label{eqn:Fvis}
|\vect{F}_\text{vis}|\approx\frac{\mu}{\sqrt{4\pi\nu\Delta t}}\lrp{\oint_{C_b}\gamma_b\mdi{l}}\equiv\rho\Gamma_b\sqrt{\frac{\nu}{4\pi\Delta t}},
\end{equation}
where $\Gamma_b$ is the net circulation established around the body. In the above, we have effectively assumed a slender body shape as the curvature of the surface was ignored. Nevertheless, this result will still serve our purpose of qualitative analysis of the viscous force. Vorticity is generated at the solid boundary by select mechanisms, namely tangential pressure gradients along the surface \citep{LighthillMJ:63a}, which includes the effect of curvature, and acceleration of the surface itself \citep{MortonBR:84a}. The appearance of these terms can be made explicit upon evaluating the Navier-Stokes equations on the boundary. In the same sense as the decomposition of \cite{GrahamWR:19a}, we can label these mechanisms as `convective' and `accelerative'. We now wish to gain a relative sense of the associated contributions to the viscous force. 

The total body circulation is divided as $\Gamma_b=\Gamma_c+\Gamma_a$, with the two terms on the right side being those due to convective and accelerative effects, respectively. Let us introduce $L$ and $U$ as length and velocity scales characteristic of the body that are appropriate measures for $\Gamma_c$. The appropriate velocity scale for $\Gamma_a$ is $\Delta U$, namely the change during time $\Delta t$ due to the acceleration. For the corresponding viscous force we must compare two characteristic acceleration scales: namely $\rho\md{U}/\md{t}\approx\rho\bp{\Delta U/\Delta t}$ and $\rho U^2/L$ for which the time scale is $T\equiv L/U$. 
Next, we define $\Gamma_c^*=\Gamma_c/UL$ and $\Gamma_a^*=\Gamma_a/\bp{\Delta UL}$, and upon substituting these into (\ref{eqn:Fvis}) and normalizing by the dynamic pressure force scaling, $\rho U^2L$, we obtain the following after a little rearrangement:
\begin{equation}
\frac{|\vect{F}_\text{vis}|}{\rho U^2 L} \hspace{3pt}\sim\hspace{3pt} \frac{1}{\sqrt{\Rey}}\sqrt{\frac{T}{\Delta t}}\lrb{\Gamma_c^*+\Gamma_a^*\frac{\Delta U}{U}}, 
\end{equation}
where $\Rey=UL/\nu$ is the Reynolds number. As expected, the overall viscous force has the classic $\Rey^{-1/2}$ scaling of boundary layer theory. We also arrive at the intuitive result that the change in body velocity must be appreciable if the vorticity/circulation generation due to acceleration is to be of the same order as that due to the instantaneous body velocity. This is, of course, in addition to the constraint $\Delta t\ll 1$ that is already implicit to the approximation (\ref{eqn:vort}).

We conclude, then, that \textit{strong} accelerations of the body provide a contribution to the vortex sheet strength on the body that is independent of that determined by its velocity, i.e. the normal boundary condition. It is important to stress the meaning of `strong' as a qualifier to the acceleration. Namely, the dynamic effect of strong vorticity generation (i.e. transport of tangential momentum) can be thought of as temporarily `stored' on the surface as a vortex sheet. The vorticity generated during a gentler acceleration will have more time to diffuse into and convect through the fluid so that $\Gamma_a$ becomes less significant to the viscous force in the context of the inviscid model.

The paper is organized as follows. The application of Kelvin's circulation theorem to allow for vorticity generation due to body acceleration is presented in \S\ref{sec:Kelvin}. The implications of infinite kinetic energy due to the circulation at infinity are discussed in \S\ref{sec:Go}. Section~\ref{sec:prob} gives the problem description of the moving body in terms of the complex potential with non-zero net circulation and entrainment on the boundaries. The generalized force calculation is presented in detail in \S\ref{sec:force}. Several examples of body  acceleration as the dominating force-generating mechanism are presented in \S\ref{sec:propel} under the context of biological propulsion. Some final concluding remarks are given in \S\ref{sec:conclude}.

\section{Kelvin's circulation theorem}\label{sec:Kelvin}
Kelvin's circulation theorem for an inviscid fluid states that the circulation $\Gamma_C$ around a \textit{closed} material curve $C$ in the fluid remains constant:
\begin{equation}
\der{\Gamma_C}{t}= 0 \quad\quad\rightarrow\quad\quad \Gamma_C(t)=\Gamma_C(t_o).
\end{equation}
More specifically, if $\Gamma_C(t_o)=0$ at some instant $t_o$ (e.g. when the fluid is initially at rest everywhere), it remains zero for all time. In applied problems, one is usually concerned with the circulation around the body and any shed vorticity, which for definiteness we give the symbol $\Gamma_o$. The typical interpretation of the theorem is that $\Gamma_o=\Gamma_C=0$. However in two dimensions, the contour $C$ technically cannot be closed around the body/shed vorticity system, but must instead extend and close elsewhere, e.g. by making a loop at infinity (see figure~\ref{fig:geom} in \S\ref{sec:force}). As such, it is possible to have $\Gamma_o\neq 0$ with the caveat that an equal circulation exists at infinity, so that when this contour segment is included as part of $C$, then $\Gamma_C=const.$ and Kelvin's theorem remains satisfied.

In the language of complex analysis (a frequent framework for modeling), Kelvin's theorem is really a restatement of Cauchy's integral theorem. The latter applies to the simply connected region defined by the `removal' of non-analytic portions of the domain. Around this contour that treats sheets of discontinuity and any other non-analytic regions as topological barriers, the net circulation is constant. We can arrive at the same conclusion by a slightly different means. Namely, Kelvin's theorem requires the assumption of a single-valued velocity field, as well as pressure, one or both of which are violated in the presence of vortex and/or entrainment sheets \citep{Mohseni:19d}. As such, the closed material contour to which the theorem applies must be a reducible curve laying entirely in the fluid domain, which is topologically equivalent to the contour in Cauchy's theorem.

The most familiar example of $\Gamma_o\neq 0$ is also perhaps the most famous and fundamental result of classical aerodynamics: the Kutta-Joukowski theory of lift for steady flow around an airfoil. The `starting vortex' is merely a physical reconciliation of the assumption $\Gamma_o=0$ with the encountered mathematics, which does not actually require nor place a starting vortex at infinity. A similar example is found in the work of \cite{Saffman:77a}, which studied the enhanced lift around a flat-plate airfoil due to a standing or `attached' leading-edge point vortex.

Note that we are \textit{not} implying that $\Gamma_o=0$ is incorrect or impossible, but rather that it is one of infinitely many permissible values that satisfy the theorem. However, we will now see the constraint that this assumption places on the representation of the physics. The generalization of Kelvin's theorem as described above can be stated as:
\begin{equation}\label{eqn:Kelvin}
\Gamma_C = \Gamma_b+\Gamma_\omega-\Gamma_\infty,
\end{equation}
where $\Gamma_b$ is the circulation around the body, $\Gamma_\omega$ is the total circulation of any free or shed vorticity, and $\Gamma_\infty$ is the circulation at infinity. The minus sign reflects the opposite direction of integration around the body and free vorticity as compared to that at infinity. Also, by definition $\Gamma_o=\Gamma_b+\Gamma_\omega$ and so we may write $\Gamma_o=\Gamma_C+\Gamma_\infty$. The vortex sheet strength on the body can be decomposed as:
\begin{equation}
\gamma_b=-\bp{\gamma_{\phi}+\gamma_{\omega}}+\gamma_{a},
\end{equation}
where $\gamma_{\phi}$ accounts for the pure potential flow around the body, $\gamma_\omega$ accounts for the flow induced at the body surface by the free vorticty, and $\gamma_a$ is the contribution from body acceleration. The first two terms are determined by satisfying the normal velocity boundary condition; the minus sign reflects the fact that this may be accomplished by an `image system'. Also, it is noted that $\gamma_\phi$ has no net circulation, $\oint\gamma_\phi\md{l}=0$, and therefore the total body circulation is:
\begin{equation}
\Gamma_b=\oint_{C_b}\gamma_b\mdi{l}=-\Gamma_\omega+\Gamma_a.
\end{equation}
Substituting this into (\ref{eqn:Kelvin}) we finally obtain:
\begin{equation}
\Gamma_o=\Gamma_a=\Gamma_C+\Gamma_\infty=\Gamma_b+\Gamma_\omega.
\end{equation}
Hence, any circulation generated by acceleration of the body must necessarily and immediately be communicated to infinity. This is regardless of the value of $\Gamma_C$, which for flows starting from rest will typically be zero. If the condition $\Gamma_o=0$ is imposed, as is conventionally done in inviscid models, we cannot possibly capture the total effect of viscous stress at the body, namely vorticity generation by acceleration of the body in addition to that created by translational motion. Next, we consider some of the physical ramifications of allowing $\Gamma_o\neq 0$.

\subsection{The circulation at infinity and kinetic energy}\label{sec:Go}
When $\Gamma_o=0$, the effect of body acceleration is relegated to the pure potential flow. In terms of the force experienced by the body, the result is the so-called `added-mass' force or `acceleration reaction', which corresponds to the inertial force on the body of finite cross-sectional area. If the body possesses geometric singularities (e.g. sharp edges), then this too can contribute to the added-mass force. This allows a mass-less body, such as the flat plate, to experience an inertial body force. We note that these forces are communicated to infinity instantaneously via the infinite signal speed of the impulse of the body surface. 

Now, if $\Gamma_o\neq 0$ then the total kinetic energy of the fluid is infinite. Despite the non-linearity of the kinetic energy, in inviscid flow the contribution to the fluid energy from added-mass forces is distinguishable from that of $\Gamma_o$ and, more notably, is finite. 
Of all the infractions on the physical world that are mathematically possible with inviscid flow, an infinite fluid energy should not give us any more pause than the others. In the absence of viscosity there is no mechanism to kill the disturbance due to the body at very large distances from it. The energy spectrum then has a contribution from infinite wave number, or zero length scale, since the flow at infinity behaves as that induced by a point singularity. Stated differently, since the flow is governed by the Laplace equation, which is elliptic, then the disturbance or impulse of any one boundary is immediately `felt' everywhere in the fluid. Hence, a non-zero $\Gamma_o$ is communicated to infinity instantaneously and results in the delivery of kinetic energy to the entire fluid at the expenditure of the work done by the boundaries, a concept which was previously described by the authors in \cite{Mohseni:18j}. This is the same mechanism that communicates the added-mass forces to infinity, albeit with a finite energy contribution as mentioned above.

Taking the boundary at infinity to be a circle of large radius $R$, then the fluid kinetic energy $T$ is logarithmically infinite as $T\approx (\Gamma_o^2/4\pi)\log(R)$ \citep{Batchelor:67a}. We can attempt to quantify the rate of change of the fluid energy due to $\Gamma_o\neq 0$ by taking the logarithmic derivative to yield:
\begin{equation}
\frac{1}{T}\der{T}{t} \hspace{3pt}\sim\hspace{3pt} 2\lrp{\frac{1}{\Gamma_o}\der{\Gamma_o}{t}}.
\end{equation}
This represents the rate at which the body must supply/lose energy to the fluid in order to accelerate through it. There is, of course, also a finite part of the energy due to the other flow components. Having motivated the physical meaning of $\Gamma_o\neq 0$, we now construct an inviscid model to include this effect.

\section{Flow due to a moving body}\label{sec:prob}
Here, we outline a generalization of the problem statement for a two-dimensional body moving within an otherwise unbounded inviscid fluid which is at rest at infinity. The phenomena of viscous diffusion and dissipation are closely linked to `entrainment', which can generally be viewed as the absorption and mixing of effectively inviscid fluid into regions of rotational fluid. The usual boundary condition of continuity of normal velocity ensures that no such entrainment will occur. This disallows the possibility of a discontinuity in the stream function. We generalize this boundary condition and model all surfaces of discontinuity as vortex-entrainment sheets that may contain mass and are dynamically distinct from the surrounding fluid \citep{Mohseni:17m,Mohseni:19d,Mohseni:20a}. The body can also be represented by such a surface of discontinuity, and as it moves these surfaces separate from the body and enter the fluid or, in physical terminology, are shed into the fluid. Since a sheet that is shed into the fluid may have a non-zero mass, then it can also support a pressure jump. Although such a sheet experiences a force, it is pursuant to its own dynamics and we shall refer to it as `free' to distinguish it from sheets that are bound to a geometry.

We begin by writing a form of the complex potential due to the motion of the body and any free sheets. The purpose is to highlight the nuances associated with entrainment rather than to obtain an expression for a specific body geometry and motion. Ultimately, we will use the model to compute the force on the body in \S\ref{sec:force}, which is complicated by the presence of net circulation and entrainment.
 
\subsection{The complex potential}\label{sec:CP}
The flow is characterized by a complex potential $W(z,t)=\phi+i\psi$, where $t$ is time, the harmonic potential is $\phi(x,y,t)$, the streamfunction is $\psi(x,y,t)$, and $z=x+iy$ is the position variable in the complex plane. The contribution $W_k$ to the total complex potential from a surface of discontinuity, labeled as $S_k$ and located at $Z_k$ in the complex plane, can be written as a Cauchy integral \citep[e.g. see][]{KelloggO:29a,MuskhelishviliNI:46a}:
\begin{eqnarray}\label{eqn:Wk}
W_k(z,t) = \frac{1}{2\pi i}\int_{S_k(t)}\frac{\mu_k(Z_k,t)\mdi{Z_k}}{z-Z_k}&=&\frac{1}{2\pi i}\int_{a(t)}^{b(t)}\pder{\mu_k^\prime}{s}\log\bp{z-Z_k(s,t)}\mdi{s}
\end{eqnarray}
where $\mu_k(s,t)$ is the \textit{complex} strength of the dipole layer representing the surface along which $s\in[a,b]$ is the arclength coordinate. The circulation and entrainment constituents of the dipole distribution are the real and imaginary parts of $\mu_k$, respectively. Accordingly, $\partial\mu_k/\partial s=\gamma_k(s,t)-iq_k(s,t)$ where $\gamma_k$ is the vortex sheet strength and $q_k$ is the entrainment sheet strength in the simple layer or vortex-entrainment sheet representation of the surface \citep{Mohseni:19d}. The total circulation $\Gamma_k$ and entrainment rate $Q_k$ associated with the sheet $S_k$ are: 
\begin{equation}\label{eqn:GQ}
\Gamma_k(t)-iQ_k(t)=\int_a^b \bp{\gamma_k-iq_k}\mdi{s} =\mu_k(b,t)-\mu_k(a,t).
\end{equation}
For a free sheet note that (\ref{eqn:Wk}) and (\ref{eqn:GQ}) imply that its endpoint in the fluid is the reference from which $\mu_k$ is measured. 

For our current problem, we write $S_\pm$ for the surfaces in the fluid that are characterized by positive and negative vorticity, respectively. Similarly, $S_b$ corresponds to the surface of the body, which will be taken as a simple closed contour. The entrainment constituent of each dipole sheet strength will be given as part of the problem set up \citep{Mohseni:20a}. For $S_\pm$, the corresponding circulation constituents are obtained as part of the calculated solution (i.e. as $S_\pm$ are shed from the body), whereas that of the body is determined from a normal boundary condition to be satisfied on the body surface $S_b$. 

By superposition the total complex potential is $W= W_{+} + W_{-} + W_b$. In what follows, we will have use for an expression of the complex potential at positions far from the body. Upon defining $\chi_k=\partial\mu_k/\partial s=\gamma_k-iq_k$ and then by applying the $|z|\rightarrow\infty$ limit to (\ref{eqn:Wk}) for each sheet we obtain:
\begin{subequations}
\begin{eqnarray}\label{eqn:Winf}
W(z,t) &\rightarrow & \frac{\Gamma_o-iQ_o}{2\pi i}\log(z)-\frac{1}{z}\lrb{\frac{1}{2\pi i}\int_{S_\pm}\chi_\pm Z_\pm\mdi{s}+\frac{1}{2\pi i}\oint_{S_b}\chi_bZ_b\mdi{s}} \\
&=& \frac{\Gamma_o-iQ_o}{2\pi i}\log(z)+\frac{1}{z}\Bb{\text{Res}\big\{W; z=\infty\big\}}
\end{eqnarray}
\end{subequations}
where $\text{Res}\big\{W;z=\infty\big\}$ is the complex residue of $W$ at infinity, $\Gamma_o=\Gamma_{+}+\Gamma_{-}+\Gamma_b$ and $Q_o=Q_{+}+Q_{-}+Q_b$. Explicit dependence on $t$ and $s$ on the right-hand side has been dropped for clarity. Also, $Z_\pm$ and $Z_b$ are the complex locations of $S_\pm$ and $S_b$, respectively. For the integrals we equally could have written $\chi_k\md{s}=\md{\mu_k}$. The net circulation around the body-sheet system is $\Gamma_o(t)$ and so is also the circulation at infinity, and similarly for $Q_o(t)$ as the net flux at infinity due to entrainment on the sheets.

In general, the vortex-entrainment sheet strength on the body $\chi_b$ will be such that the integral over $S_b$ in (\ref{eqn:Winf}) gives rise to $O(z^{-1})$ terms that are proportional to the body motion and geometry. More specifically, translation of the body results in a non-vanishing dipolar contribution to the complex potential; see \cite{LlewellynSmith:08a} for an excellent work regarding such dipole moments. The terms related to the motion and geometry obviously provide direct contributions to the force on the body. The generalized force calculation in the presence of a net circulation as well as entrainment and/or sheets with mass is discussed next in \S\ref{sec:force}. 

\section{Force calculation}\label{sec:force}
As mentioned in \S\ref{sec:intro}, the representation of viscous shear forces in an inviscid model can be accomplished with singularities whose induced flows alter the pressure on the body. As such, the net force on the body is still computed as the integral of fluid pressure. Owing to the possibility of entrainment on the body surface and of non-zero mass on the free sheets in the fluid, the usual force calculation process by the Blasius integral must be altered. The pressure force on the body (ignoring gravitational effects) is:
\begin{equation}\label{eqn:Fb}
F_x+iF_y=\oint_{C_b} ip\mdi{z}=-i\rho\oint_{C_b}\lrb{\pder{\phi}{t}+\frac{1}{2}|\grad{\phi}|^2}\mdi{z},
\end{equation}
where $C_b$ is a contour in the fluid but adjacent to the body surface. The fluid velocity components tangential and normal to the contour are $u_s$ and $u_n$, and similarly for the corresponding components of the velocity of the contour, $v_s$ and $v_n$. Also, in (\ref{eqn:Fb}) we have omitted a reference pressure $p_o(t)$ as it has no net contribution to the force. Momentarily borrowing vector notation for $\vect{v}=(v_s,v_n)$ and $\vect{u}=\grad{\phi}=(u_s,u_n)$, we have:
\begin{equation}\label{eqn:phi}
\der{\phi}{t}=\pder{\phi}{t}+\vect{v}\cdot\grad{\phi}=\pder{\phi}{t}+\bp{v_su_s+v_nu_n}.
\end{equation}
\cite{SedovLI:65a} substitutes (\ref{eqn:phi}) into (\ref{eqn:Fb}) and then considers the force on a portion $\md{z}$ of the contour by expanding the total derivative of the harmonic potential. Upon performing some further manipulations, he then employs the assumption of a no through-flow boundary condition on $C_b$ to obtain his final expression. This assumption eliminates the appearance of a term containing $(\md{z}/\md{t})\md{\psi}$. However, for a non-zero flux on $C_b$ this term represents the rate at which mass is entrained into the body surface sheet $S_b$ (i.e. the boundary layers). Hence, with $\md{z}/\md{t}$ as the velocity of the surface, this term quite literally represents added mass. The other terms give rise to force contributions from net and shed circulations, changes in body shape, and hydrodynamic impulses of the boundaries.

Alternatively, we may use Cauchy's integral theorem
\begin{equation}\label{eqn:Cauchy}
\oint_C f(z)\mdi{z}=0
\end{equation}
for a function $f(z)$ that is analytic within the simply connected fluid region $D_f$ bounded by the \textit{closed} contour $C=\partial D_f$. However, if we remain working with $\phi$, then the resulting force expression does not clearly elucidate the effect of the body, most especially the entrainment on its surface. Moreover, while the expression can be written in terms of the residue of the complex potential at infinity, one cannot arrive at the result as directly as in the case of the Blasius integral where the method of residues is valid. Specifically, since $\phi=\text{Re}\{W\}$, then the result is a `partial residue' at infinity with a $1/2$ factor appearing in front of the usual complex residue result. It is easy to verify that the `other half' of the complex residue from a dipolar distribution comes the imaginary part $\psi=\text{Im}\{W\}$, that is the stream function. 

\begin{figure}
\begin{center}
\begin{minipage}{0.5\linewidth}\begin{center}
\includegraphics[width=0.99\textwidth, angle=0]{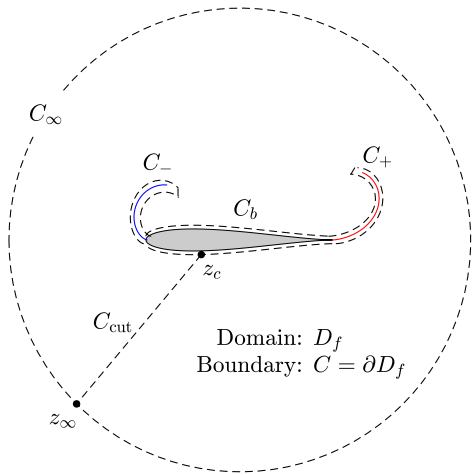}
\end{center}\end{minipage}
\caption{Schematic of the geometry and definitions of the simply connected fluid domain $D_f$ and its boundary $C =\partial D_f$ (dashed lines). The outer boundary is $C_o=C_\infty\cup C_\text{cut}$ where $C_\infty$ encloses the inner boundary $C_i=C_b\cup C_{+}\cup C_{-}$ and the two are connected by the branch cut $C_\text{cut}$. The intersections of $C_\text{cut}$ at the body and at infinity are $z_c$ and $z_\infty$, respectively. }
\label{fig:geom}
\end{center}
\end{figure}
Fortunately, there is a judicious combination of Sedov's approach and the Blasius integral that obviates the effects of entrainment and allows use of the method of residues. Additionally, the expression easily reduces to the conventional result when all entrainment terms are set to zero. To start, we replace the harmonic potential with $\phi=W-i\psi$ in (\ref{eqn:phi}) and upon substituting the result into (\ref{eqn:Fb}) we obtain
\begin{equation}\label{eqn:Fb2}
F_x+iF_y = -\rho\oint_{C_b}\der{\psi}{t}\mdi{z}-i\rho\oint_{C_b}\lrb{\der{W}{t}+\tfrac{1}{2}(u_s^2+u_n^2)-(v_su_s+v_nu_n)}\mdi{z}.
\end{equation}
The first integral will be directly computed on $C_b$, while we will apply Cauchy's integral theorem to the second integral to convert the contour to integrals over the other parts of the fluid boundary. Let us begin with this second integral as it is more familiar to the conventional problem. The total fluid boundary $C=\partial D_f$ is divided up as $C=C_o\cup C_i$, where $C_o=C_\infty\cup C_\text{cut}$ and $C_i=C_b\cup C_{+}\cup C_{-}$ as seen in figure~\ref{fig:geom}. Hence, $C_i$ is an `inner' contour immediately surrounding the body-sheet system and $C_o$ is an `outer' contour enclosing $C_i$ via $C_\infty$ and connected to it by a branch cut $C_\text{cut}$. 

Next, we take $C_\infty$ as a circular contour of indefinitely large radius $R$, and let $z_c$ and $z_\infty=R\e{i\theta_\infty}$ be the locations on $C_b$ and $C_\infty$, respectively, where the cut $C_\text{cut}$ intersects (again see figure~\ref{fig:geom}). By continuity of the velocity along $C_\text{cut}$ and the absence of a \textit{complex} residue in the individual velocity components at infinity, the integrals of the velocity terms in (\ref{eqn:Fb2}) over $C_o$ are zero. The remaining integral must be computed with coordinates having an origin somewhere inside $C_i$ so that the cyclic part of the complex potential, i.e. $-\jb{W}=\Gamma_o-iQ_o$, is `looped' around. Let $\tilde{z}_c$ and $\tilde{z}_\infty$ be the corresponding positions in this axis system. Carefully evaluating the integral by using (\ref{eqn:Winf},\textit{b}) on $C_\infty$ and the fact that the jump in $W$ across the cut does not vary along $C_\text{cut}$ then gives:
\begin{eqnarray}
-i\rho\oint_{C_o}\der{W}{t}\md{z}&=&-i\rho\underbrace{\der{}{t}\Bb{\bp{\Gamma_o-iQ_o}\bp{\tilde{z}_\infty-\tilde{z}_c}}}_{C_\text{cut}} \nonumber \\
&&-i\rho\underbrace{\der{}{t}\Bb{-\bp{\Gamma_o-iQ_o}\tilde{z}_\infty+2\pi i\text{Res}\big\{W; z=\infty\big\}}}_{C_\infty} \label{eqn:Finf}\\
&=&i\rho\der{}{t}\Bb{\bp{\Gamma_o-iQ_o}\tilde{z}_c}+i\rho\der{}{t}\lrb{\int_{S_\pm}\chi_\pm Z_\pm\mdi{s}+\oint_{S_b}\chi_bZ_b\mdi{s}}, \label{eqn:Finf2}
\end{eqnarray}
independent of the radius $R$ and $S_\pm=S_{+}\cup S_{-}$ implies summation. Now we turn to the contours $C_\pm$ that surround the sheets $S_\pm$ in the fluid. Since they are infinitely thin, then the contour integral can be interchanged for a line integral along the sheet. Evaluation of the integrand in (\ref{eqn:Fb2}) on the sheets then yields:
\begin{equation}\label{eqn:Fpm}
\rho\int_{S_\pm}\lrb{\der{}{t}\bp{\Gamma-iQ}+(\overline{u_s}-v_s)\gamma+(\overline{u_n}-v_n)q}i\e{i\theta}\mdi{s},
\end{equation}
where it is understood that $\Gamma=-\jb{\phi}$ and $Q=\jb{\psi}$ are the circulation and entrainment rate up to a given arclength position $s$ on the sheet. Also, $\overline{u_s}$ and $\overline{u_n}$ are the averages of the fluid velocity component jumps (i.e. principal values), and $\theta(s,t)$ is the local argument of the sheet relative to the horizontal so that $i\e{i\theta}=\uv{n}$ is the varying normal vector.

Now, to compute the first integral in (\ref{eqn:Fb2}) over $C_b$ we write:
\begin{equation}\label{eqn:dF}
-\der{\psi}{t}\md{z}=-\md{\lrp{z\der{\psi}{t}}}+\der{}{t}\bp{z\md{\psi}}-\der{z}{t}\md{\psi}.
\end{equation}
Since the contour is on the body, the coordinate system must similarly have origin somewhere within $C_b$ to `loop around' the cyclic part of $\psi$. Letting the coordinate variable again be $\tilde{z}$, the first term is a total differential and is easily integrated to each side of the discontinuity of $\psi$ located at $\tilde{z}_c$, which gives $\tilde{z}_c\md{\jb{\psi}}/\md{t}=\tilde{z}_c\md{Q_b}/\,d{t}$. Following Sedov, the second term integrates to:
\begin{equation}
\oint_{C_b}\der{}{t}\bp{z\md{\psi}}=\der{^2}{t^2}\bp{z_b A_b},
\end{equation}
where $A_b$ is the cross-sectional area of the body and $z_b$ is its center of gravity. For the third term, we recognize that the contour velocity $\md{z}/\md{t}$ is equal to the body velocity, which we write as $(U_{s,b}+iU_{n,b})\e{i\theta_b}$ with $U_{s,b}$ and $U_{n,b}$ as the components tangential and normal to the contour whose argument relative to the horizontal is $\theta_b$. We then have:
\begin{eqnarray}\label{eqn:Qb}
-\oint_{C_b}\der{z}{t}\mdi{\psi} = \oint_{C_b}\bp{U_{s,b}+iU_{n,b}}u_{n,f}\mdi{z} 
\end{eqnarray}
where we have used $\md{\psi}=\bp{\partial\psi/\partial s}\md{s}=-u_{n,f}\md{s}$ and $\e{i\theta_b}\md{s}=\md{z}$ with $u_{n,f}$ as the \textit{fluid} velocity adjacent to the body. On account of the entrainment boundary condition we have $u_{n,f}=U_{n,b}+q_b$. For a translational component of the body velocity, say $U_b+iV_b$, the above gives the simple result $(U_b+iV_b)Q_b$, which can be obtained from either the left- or right-hand side upon using $-\oint\md{\psi}=Q_b$ or $\oint q_b\md{s}=Q_b$.

Next, without loss of generality we may take the $\tilde{z}$ origin to be fixed in the body, which translates with velocity $U_b+iV_b$. As such:
\begin{equation}\label{eqn:dzc}
\der{\tilde{z}_c}{t}=\lrp{\der{\tilde{z}_c}{t}}_b+(U_b+iV_b)=u_c\e{i\theta_c}+(U_b+iV_b), 
\end{equation}
where $\bp{\cdot}_b$ indicates the body-fixed frame and $u_c(s,t)$ is the velocity of $\tilde{z}_c$ relative to the body and is tangential to its surface $S_b$ so that $\e{i\theta_c}=\uv{s}$ is the tangent vector at $\tilde{z}_c$. 

Collecting (\ref{eqn:Finf})--(\ref{eqn:Qb}) and upon using (\ref{eqn:dzc}) we finally obtain the total force exerted on the body as:
\begin{eqnarray}\label{eqn:F}
F_x+iF_y&=&-i\rho\der{}{t}\Bb{2\pi i\text{Res}\big\{W,z=\infty\big\}}+\rho\der{^2}{t^2}\bp{z_bA_b} \nonumber \\ 
&&+i\rho\bp{\Gamma_o-iQ_o}\bp{U_b+iV_b}+i\rho\lrp{\der{}{t}\Bb{\bp{\Gamma_o-iQ_o}\tilde{z}_c}}_b \nonumber \\
&&+\rho\int_{S_\pm}\left[\der{}{t}\big(\Gamma-iQ\big)+(\overline{u_s}-v_s)\gamma+(\overline{u_n}-v_n)q\right]i\e{i\theta}\mdi{s} \nonumber \\
&&+\rho\tilde{z}_c\der{Q_b}{t}+\bp{U_b+iV_b}Q_b+\oint_{C_b}\bp{U_{s,b}^\prime+iU_{n,b}^\prime}\bp{U_{n,b}+q_b}\mdi{z} .
\end{eqnarray}
Each term in (\ref{eqn:F}) is, in one form or another, the rate-of-change of the hydrodynamic impulse of the boundaries of the fluid domain $D_f$. The first line is the conventional result when there is zero entrainment everywhere and no mass on the boundaries. This includes the `vortex force' and the inertial force on the body when it accelerates or changes its size and shape. The second line has the forces due to net circulation and entrainment, including the familiar Joukowski force and an analogous force due to the movement of the cut intersection $\tilde{z}_c$ on the body. 
The integrals over $S_\pm$ on the third line represent the effect of acceleration of the mass contained in the free sheets. In this regard, these sheets are similar to flexible membranes with different mass density than the fluid. For a recent study on extensible membrane flutter see \cite{AlbenS:20a}. The fourth line has forces due to entrainment specifically at the body, where $U_{s,b}^\prime$ and $U_{n,b}^\prime$ are the non-translational components of the body velocity (e.g. rotation, deformation).
 

At this stage $\Gamma_o(t)$ and $\tilde{z}_c(t)$ remain unspecified, and both must be given to render the problem closed and unique. The velocity field is determined with $\Gamma_o$, whereas knowledge of the cut intersection position $\tilde{z}_c$ is only required to make the force unique. We note that for a steady problem or an unsteady one in which $\Gamma_o\equiv 0$, the specification of the cut is arbitrary in that it has no bearing on the velocity field or force on the body. However, it is still technically required to make the complex potential unique. Hence, for the general case the uniqueness of the force follows from uniqueness of the complex potential. Next, we consider some simple examples to elucidate the importance of these two quantities.

\section{Wake-less self-propulsion of a body}\label{sec:propel}
Here, we demonstrate how the specification of $\Gamma_o(t)$ and $\tilde{z}_c(t)$ can generate a physically meaningful force. Recall that these quantities are the net circulation established by the acceleration of the body and the intersection of the branch cut on the body. We consider very simple solutions representing the self-propulsion of a body by some oscillatory mechanism. The usual approach to such a problem is the intuitive one that models the shedding and subsequent convection of vorticity into the wake. If this task has been completed under the assumption $\Gamma_o=0$, then one could evaluate the vortex force due to the wake vorticity. As an alternative approach, we neglect the details of the wake and instead claim to have knowledge of the \textit{moving} stagnation point on the body surface, which is a natural choice for the branch point $\tilde{z}_c$. The motivation for this simplification is the observation that biological propulsion often involves very large accelerations and displacements of the animal's body and/or appendages. In such cases, the propulsive forces ought to be dominated by the vorticity generation due to the acceleration, with the details of the induced effects of the wake being of lesser importance. 

The body is prescribed as \textit{steadily} translating with velocity $U_b+iV_b=-U$, i.e. to the left, and with some circulation $\Gamma_o(t)$. As such, a negative value of the circulation results in a positive side force or lift, and for convenience we let $\Gamma_o\rightarrow -\Gamma_o$. Further setting the entrainment to zero, then the only surviving force terms from (\ref{eqn:F}) are:
\begin{equation}
F_x+iF_y=i\rho\lrb{\Gamma_oU-\der{}{t}\big(\Gamma_o\tilde{z}_c\big)_b}.
\end{equation}
The quantity $\rho\Gamma_o\tilde{z}_c$ is the net pressure impulse on the branch point $\tilde{z}_c$, which is part of the boundary. In this way we can interpret the point as an `appendage' of the body that pushes on the fluid to create propulsive forces (see figure~\ref{fig:bird}(\textit{b})). Our objective now is to prescribe $\Gamma_o(t)$ and $\tilde{z}_c(t)$ and interpret the resulting forces.

For simplicity, we take the body to be a circular cylinder of radius $a$ so that $\tilde{z}_c=a\e{i\theta_c}$. However, the following results are extendable to any simply connected body via the topological equivalence of a Jordan curve to the cylinder surface. For intuitive purposes, we let the lift be the side force, $L=F_y$, and the thrust be the negative of the axial force, $T=-F_x$, and therefore:
\begin{subequations}\label{eqn:cyl_F}
\begin{eqnarray}
T &=& -\rho \lrb{a\sin\theta_c\der{\Gamma_o}{t}+a\Gamma_o\cos\theta_c\der{\theta_c}{t}} \\
L &=& \rho\lrb{\Gamma_oU-a\cos\theta_c\der{\Gamma_o}{t}+a\Gamma_o\sin\theta_c\der{\theta_c}{t}}.
\end{eqnarray}
\end{subequations}
Assuming the flow to be oscillatory, we describe the location of the stagnation point as $\theta_c=\theta_o\sin(\Omega t-\alpha)$ where $\theta_o$ is the amplitude, $\Omega$ the angular frequency and $\alpha$ is a phase shift. Therefore, $\theta_c$ oscillates about the horizontal axis at the rear of the body (see figure~\ref{fig:bird}(\textit{b})). It is simpler to work in the state-space with $\theta_c$ as the independent variable rather than time $t$. As such,
\begin{eqnarray}
\der{\theta_c}{t}=\Omega\theta_o\sqrt{1-\left(\frac{\theta_c}{\theta_o}\right)^2}
\end{eqnarray}
and we note that $\Omega$ must now be treated as a signed quantity, i.e. as the angular velocity, to reflect the oscillation of $\theta_c$. The following subsections give examples of different propulsive mechanisms. Each subsequent example builds on the previous one by introducing a further degree of complexity.

\subsection{Flying bird}\label{sec:bird}
\begin{figure}
\begin{center}
\begin{minipage}{0.49\linewidth}\begin{center}
\includegraphics[width=0.99\textwidth, angle=0]{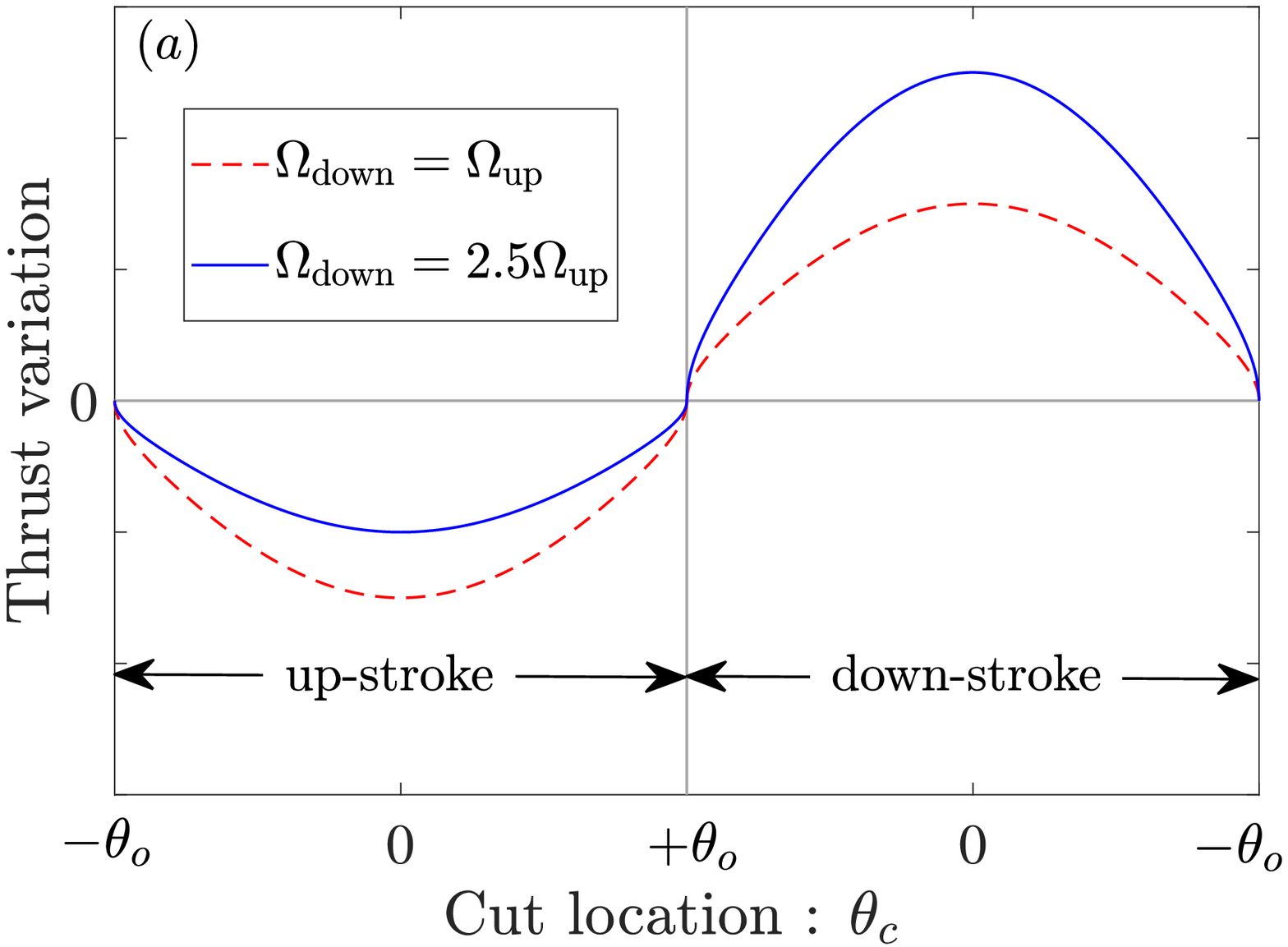}
\end{center}\end{minipage}
\begin{minipage}{0.49\linewidth}\begin{center}
\includegraphics[width=0.99\textwidth, angle=0]{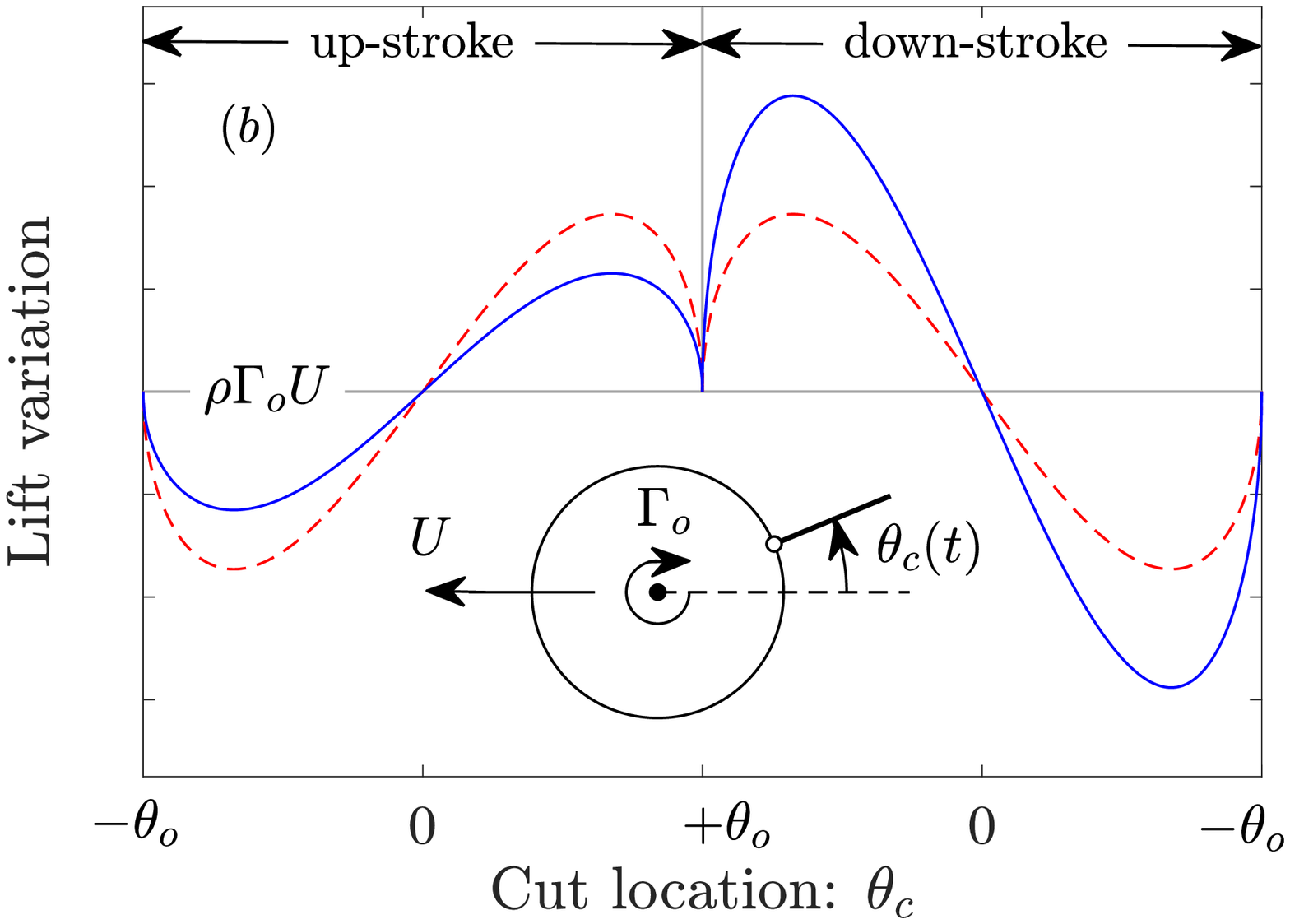}
\end{center}\end{minipage}
\caption{Force components for the self-propelling cylinder with an oscillating branch cut stagnation point. These plots correspond to a constant $\Gamma_o$ and an amplitude $\theta_o=\pi/4$. The legend applies to both plots. Two cases are shown: equal up-down strokes (dashed red) and asymmetric up-down strokes (solid blue). (\textit{a}) Thrust over a cycle of the motion and (\textit{b}) lift or side force over the same cycle. The inset is a schematic of the geometry.}
\label{fig:bird}
\end{center}
\end{figure}
Let us first consider the simple case of a constant circulation. Figure~\ref{fig:bird}(\textit{a}) plots an example of the thrust variation as $\theta_c$ travels through one cycle, defined as movement from $(-\theta_o)$ to $(+\theta_o)$ and back again. The body experiences a zero-net axial force with a drag on the first half-cycle (up-stroke), and a thrust on the second half-cycle (down-stroke). This is similar to the flapping of a bird wing. Indeed, the oscillating lift component plotted in figure~\ref{fig:bird}(\textit{b}) shows that $L$ is an odd function over each half-cycle with negative lift occurring on the initial part of the up-stroke. This becomes positive when the `wing' passes the horizontal and begins decelerating so that there is a positive pressure difference across it caused by the impinging inertia of the fluid below the wing. This phenomenon is sometimes referred to as `wake capture' \citep{Dickinson:03a,SaneSP:03a}. The reverse process occurs on the down-stroke and the oscillating lift component also has a zero-net contribution over the full cycle. As such, the body propels itself forward with constant velocity $U$ by oscillations of the branch-point appendage at its rear. It also maintains a steady Joukowski lift $\overline{L}=\rho U\Gamma_o$ since $\Gamma_o$ is a non-zero constant. As such, the problem geometry can be loosely interpreted as the side view of a flying bird cruising at constant altitude.

These results, while interesting, are not particularly realistic. The thrust experienced on the down-stroke is exactly negated by the drag on the up-stroke. A more feasible result can be obtained by noting that animals often exhibit asymmetry in their propulsive motions. In our current bird-like motion, the down-stroke is comparatively faster. Let $\Omega_\text{up}$ be the angular velocity for the up-stroke and $-\Omega_\text{down}$ be that of the down-stroke. Example thrust and lift traces with $\Omega_\text{down}=2.5\Omega_\text{up}$ are also shown in figures~\ref{fig:bird}(\textit{a}) and~(\textit{b}). The oscillating lift component remains an odd function over the half-cycle and so maintains a zero-net contribution. The average thrust over a full cycle ($4\theta_o$) is:
\begin{equation}
\frac{\overline{T}}{\rho U \Gamma_o}=\frac{\pi}{2}\bp{k\theta_o}J_1(\theta_o)\lrb{1-\frac{\Omega_\text{up}}{\Omega_\text{down}}},
\end{equation}
where $k=a\Omega_\text{down}/2U$ is the reduced frequency based on the propulsive down-stroke and $J_1(x)$ is the first-order Bessel function of the first kind. For a given set of parameters the average thrust is maximum at about $\theta_o=137^o$. Next, we present an example that is representative of the physics of fish locomotion.

\subsection{Swimming fish}\label{sec:fish}
Here, a representation of a swimming fish can be obtained by prescribing an oscillatory circulation generated by the `caudal fin' branch point as it flaps. We assume the fish is neutrally-buoyant and thus continue under the auspice that it generates zero-net side force. We investigate two sub-cases, which we will argue correspond to a fast, Carangiform-type locomotion and a smooth undulatory-type locomotion. 

\subsubsection{Asymmetric flapping}\label{sec:asym}
We consider the fast case first and employ the asymmetry in the angular speed of the flapping motion as in \S\ref{sec:bird}. Here, however, the asymmetry occurs every quarter-cycle. Namely, the fish flaps its fin faster during the return strokes toward the center line, i.e. from $\pm\theta_o$ to $0$, to push itself forward and flaps slower on advancing strokes away from $\theta_c=0$ to lessen the drag penalty. Moreover, we assume this fast motion to be so aggressive that the strongest circulation is generated at the stroke reversals when the fin is at its maximum displacement amplitude. This is due to a significant wake capture effect \citep[see also][]{DeVoriaAC:13b}. To represent this, we write $\Gamma_o(\theta_c)=\Gamma_v\sin\bp{\pi\theta_c/2\theta_o}$ where $\Gamma_v$ is the circulation magnitude of the vortex generated by the fin. Since the acceleration of the `fin' is $a\bp{\md{^2\theta_c}/\md{t^2}}=-a\Omega^2\theta_c$, then $\Gamma_o$ is out of phase with the acceleration (their magnitudes are in-phase, though). As we will see, this circulation production corresponds to a reverse von K{\'a}rm{\'a}n vortex street, namely a propulsive wake configuration. With this form the forces can be written as:
\begin{subequations}\label{eqn:fish_force}
\begin{eqnarray}
T &=& -\rho a\Gamma_v\der{\theta_c}{t}\lrb{\frac{\pi}{2\theta_o}\sin\theta_c\cos\lrp{\frac{\pi\theta_c}{2\theta_o}}+\cos\theta_c\sin\lrp{\frac{\pi\theta_c}{2\theta_o}}} \\
F_y &=& \rho\Gamma_vU\sin\lrp{\frac{\pi\theta_c}{2\theta_o}}-\rho a\Gamma_v\der{\theta_c}{t}\lrb{\frac{\pi}{2\theta_o}\cos\theta_c\cos\lrp{\frac{\pi\theta_c}{2\theta_o}}-\sin\theta_c\sin\lrp{\frac{\pi\theta_c}{2\theta_o}}}
\end{eqnarray}
\end{subequations}
where we have returned to $F_y$ as the side force given the context of the bio-propulsive mechanism. The first term in the side force is an odd function over the half-cycle and is not effected by the asymmetry in the angular velocity, and so it has zero-net contribution. The second term is an even function over the half-cycle, but since the asymmetry occurs every quarter-cycle this term also yields no net side force. As such, the side force is an odd function over the full cycle so that $\overline{F_y}=0$ as we required. 

\begin{figure}
\begin{center}
\begin{minipage}{0.49\linewidth}\begin{center}
\includegraphics[width=0.99\textwidth, angle=0]{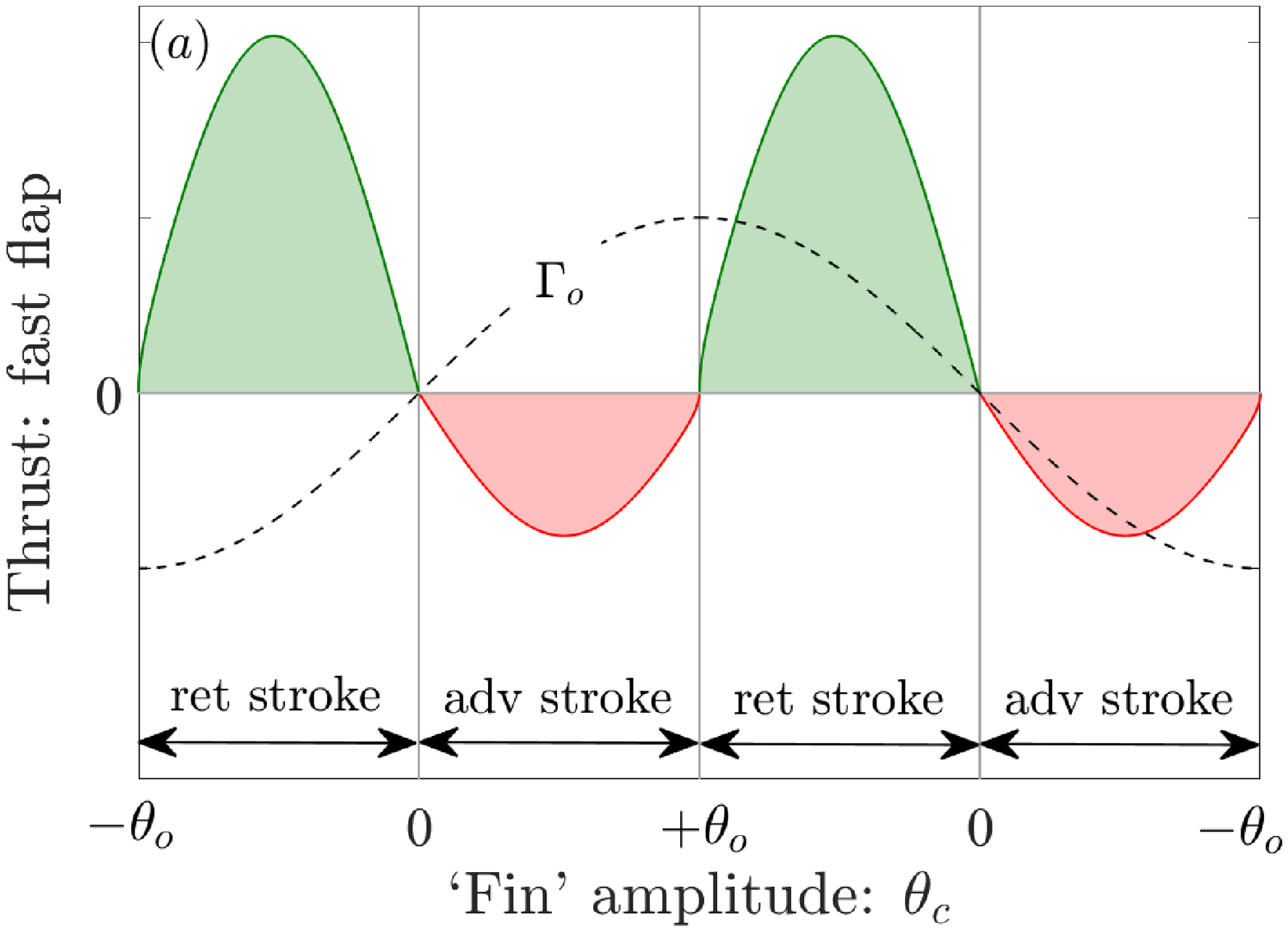}
\end{center}\end{minipage}
\begin{minipage}{0.49\linewidth}\begin{center}
\includegraphics[width=0.99\textwidth, angle=0]{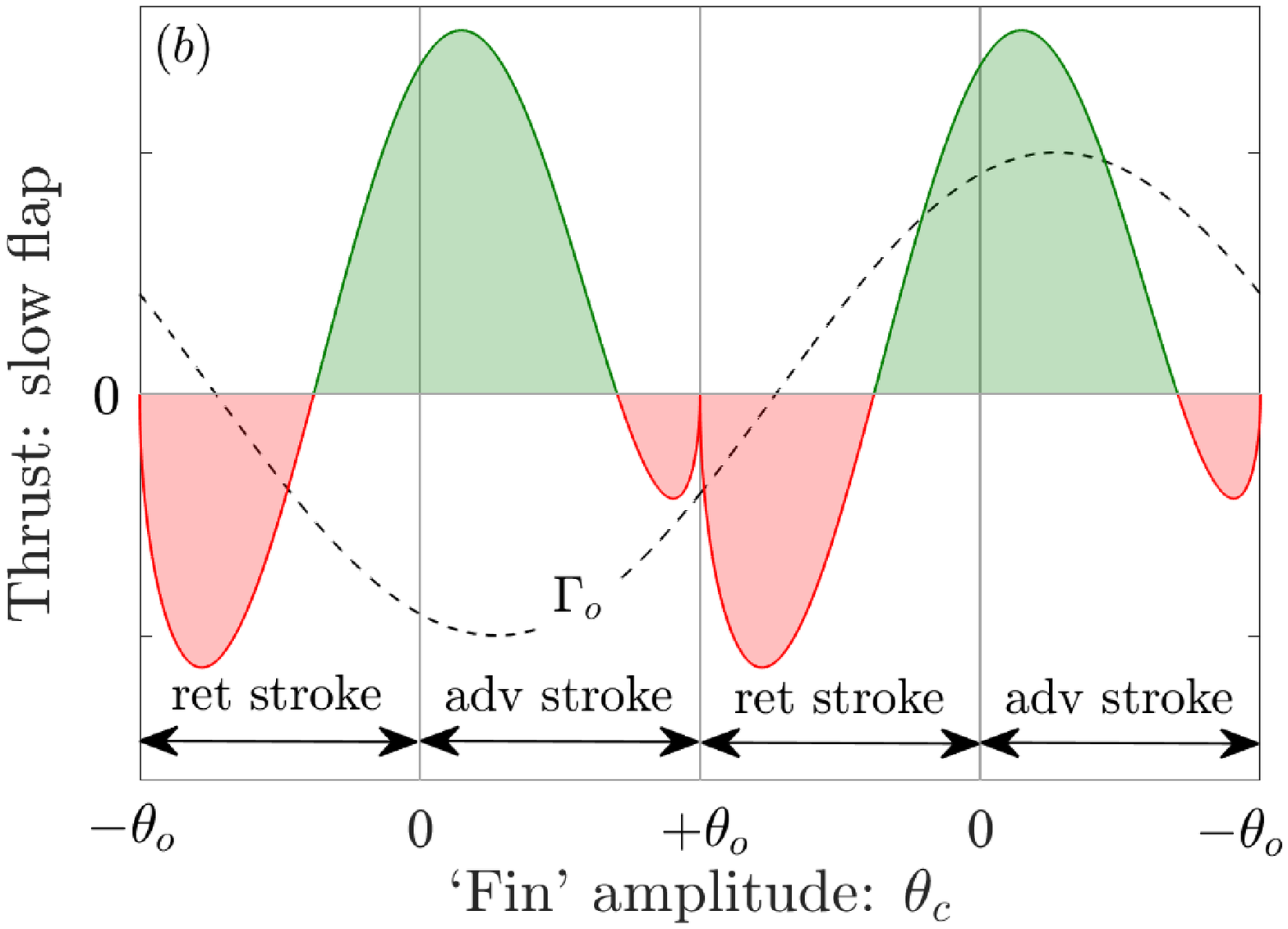}
\end{center}\end{minipage}
\caption{Modeled thrust by a `fish caudal fin' in different propulsive modes. (\textit{a}) Fast, asymmetric Carangiform-type locomotion with high acceleration. There is asymmetry in the angular speed of the returning and advancing strokes as $\Omega_\text{ret}=2.5\Omega_\text{adv}$.  (\textit{b}) Slow, symmetric undulatory-type locomotion with reduced frequency $k=1$. The asymmetry is caused by a phase shift in circulation generation. For each case $\theta_o=\pi/4$ and the circulation production $\Gamma_o(\theta_c)$ is shown as the dashed black line. The simplified geometry is the same as in figure~\ref{fig:bird}.}
\label{fig:fish}
\end{center}
\end{figure}
Figure~\ref{fig:fish}(\textit{a}) plots an example of the thrust variation due to the asymmetric fin oscillation yielding a positive net result. Here the wake capture effect is strong as the fin quickly snaps back at the stroke reversal and `pushes' off of the wake it just created on the advancing stroke with comparatively lower speed. Also plotted is the circulation $\Gamma_o(\theta_c)$ and, by design, takes its greatest values $\pm\Gamma_v$ at the stroke reversals $\pm\theta_o$. As mentioned previously, this is a propulsive wake configuration representing a reverse von K{\'a}rm{\'a}n vortex street, which would have some lateral spacing of the vortices determined partly by the oscillation amplitude. This propulsive mode undoubtedly requires a high energy expenditure as the thrust and fin velocity are such that the fish does work on the fluid over the entire cycle.

\subsubsection{Symmetric flapping}\label{sec:sym}
Next, we consider the second case of a less aggressive locomotion. During steady `cruising' fish often exhibit smooth sinuous motions of the tail while still propelling forward. Therefore, we take the angular speeds of the strokes $\Omega$ to be the same, i.e. symmetry of the oscillation profile. In this case we expect that the circulation production and the wake capture effect may not be exactly out of phase with the acceleration. Moreover, for very long periods of oscillation the fin would generate a vortex whose strength increases from zero at the maximum displacement amplitude. These features are represented by a circulation production of the form $\Gamma_o(t)=-\Gamma_v\cos\bp{\pi\theta_c/2\theta_o-\beta}$. Here, $\Gamma_v$ is to be treated as a signed quantity to reflect the corresponding change in vorticity from the first half-cycle to the second, as we similarly did with $\Omega$ for the stroke reversal. As such, $\beta$ corresponds to a phase shift of $\Gamma_o$ from the velocity, where $\beta=0$ is exactly out of phase (i.e. positive angular velocity creates negative circulation and vice versa). The thrust and side force equations are very similar to (\ref{eqn:fish_force}) and can be obtained from them by replacing $\cos\bp{\pi\theta_c/2\theta_o}$ with $\sin\bp{\pi\theta_c/2\theta_o-\beta}$ and $\sin\bp{\pi\theta_c/2\theta_o}$ with $-\cos\bp{\pi\theta_c/2\theta_o-\beta}$. In the resulting equations it should be noted that the product $\Gamma_v\md{\theta_c}/\md{t}$ is now always positive.

Now recall our requirement that no net side force is generated. Even with a non-zero phase shift $\beta\neq 0$, the side force is an odd function about the \textit{full} cycle. While this satisfies our condition, we note that if the motion is a slower undulation, then it is physically plausible to have zero-net side force on the \textit{half}-cycle. In other words, for a case with moderate acceleration we don't expect the side force to be appreciable on any part of the cycle. Hence, we can additionally impose a zero-net side force condition on the half-cycle, which determines the necessary phase shift $\beta$ for a given amplitude $\theta_o$ and reduced frequency $k=a\Omega/2U$ as:
\begin{subequations}
\begin{eqnarray}\label{eqn:beta} 
\frac{2\cos\beta}{k\pi} &=& -\theta_o\int_{-1}^{1}\sqrt{1-\xi^2}F_y^*(\xi)\mdi{\xi} \\
F_y^*(\xi) &=& \frac{\pi}{2\theta_o}\cos\bp{\theta_o\xi}\sin\lrp{\frac{\pi\xi}{2}-\beta}+\sin\bp{\theta_o\xi}\cos\lrp{\frac{\pi\xi}{2}-\beta}.
\end{eqnarray}
\end{subequations}
Figure~\ref{fig:fish}(\textit{b}) shows the modeled thrust generation of a slow flap for the case of $\theta_o=\pi/4$ and $k=1$, which yields a circulation phase \textit{lag} of $\beta\approx 24^o$ as determined from the condition of zero-net side force on the half-cycle in (\ref{eqn:beta}); the integral was computed numerically. Here, the wake capture effect is used both actively and passively. Namely, at the beginning of the first return stroke the fin is moving slower relative to the forward motion of the fish and thus experiences a pressure drag. As the fin picks up speed toward the symmetry plane it begins to actively `push' on the wake as in the fast acceleration case of figure~\ref{fig:fish}(\textit{a}). Beyond the center line, the decelerating fin passively `coasts' forward on the wake it created during the first return stroke. As it slows down again near the stroke reversal, a similar pressure drag is incurred. However, this advancing-stroke drag penalty is lower than that in the return stroke because the wake is still impinging on the fin and counteracts the dynamic pressure due to the forward motion of the fish. At the stroke reversal the cycle repeats.

\subsubsection{Net thrust production of locomotive modes}\label{sec:thrust}
It is worthwhile to investigate the stroke-averaged thrust and its dependency on the amplitude and reduced frequency (or the closely related Strouhal number $St$). For the general case with some given $\Gamma_o(\theta_c)=\Gamma_vG\bp{\theta_c}$, we can write the thrust as:
\begin{subequations}
\begin{eqnarray}
T(\theta_c) &=& \rho a\Omega\Gamma_v\theta_o\Bb{-\sqrt{1-\xi^2}\hspace{4pt}T^*(\xi)} \\
T^*\bp{\xi} &=& \sin\bp{\theta_o\xi}\der{G}{\theta_c}+\cos\bp{\theta_o\xi}G
\end{eqnarray}
\end{subequations}
where $\xi=\theta_c/\theta_o$ and $T^*$ represents the thrust variation in the normalized variable $\xi$. For the asymmetric flapping case, whose asymmetry is defined on the quarter-cycle, the average thrust over a full cycle is:
\begin{eqnarray}
\frac{\overline{T}_a}{\rho U\Gamma_v} &=& k\theta_o\lrp{\int_0^1\sqrt{1-\xi^2}\hspace{4pt}T_a^*(\xi)\mdi{\xi}}\lrb{1-\frac{\Omega_\text{adv}}{\Omega_\text{ret}}} \nonumber \\
& \equiv & k\theta_oI_a\lrb{1-\frac{\Omega_\text{adv}}{\Omega_\text{ret}}}. \label{eqn:Tfast} 
\end{eqnarray}
where again $k=a\Omega_\text{ret}/2U$ is the reduced frequency of the propulsive stroke. For the symmetric flapping case, we obtain a similar result but the asymmetry is over the half-cycle so that the integral in $\xi$-space is from $-1$ to $+1$:
\begin{eqnarray}
\frac{\overline{T}_s}{\rho U\Gamma_v} &=& k\theta_o\lrp{\int_{-1}^{1}\sqrt{1-\xi^2}\hspace{4pt}T_s^*(\xi)\mdi{\xi}} \nonumber \\
& \equiv & k\theta_oI_s \label{eqn:Tmod} 
\end{eqnarray}
where $k=a\Omega/2U$ is the reduced frequency of the symmetric flapping motion. In each flapping case the normalized average thrust is linearly proportional to the reduced frequency. However, we have not yet considered how $\Gamma_v$ varies with $k$, which will be required in order to make quantitative comparison to experiments, specifically those on oscillating foils of chord length $c$ and symmetric flapping frequency $\Omega$. To this end, we briefly address the non-dimensionalization. Typically, force coefficients (per-unit-depth) are defined using the dynamic pressure and a length scale for normalization. In the experiments the thrust coefficient is $C_T=2\overline{T}/\rho U^2 c$ and likewise the reduced frequency is $k=2\pi f c/2U$ where $f=\Omega/2\pi$ is the foil oscillation frequency in Hertz. By using these definitions, we are essentially just making the \textit{diameter} of the cylinder the relevant length scale: $2a=c$. 

\begin{figure}
\begin{center}
\begin{minipage}{0.49\linewidth}\begin{center}
\includegraphics[width=0.99\textwidth, angle=0]{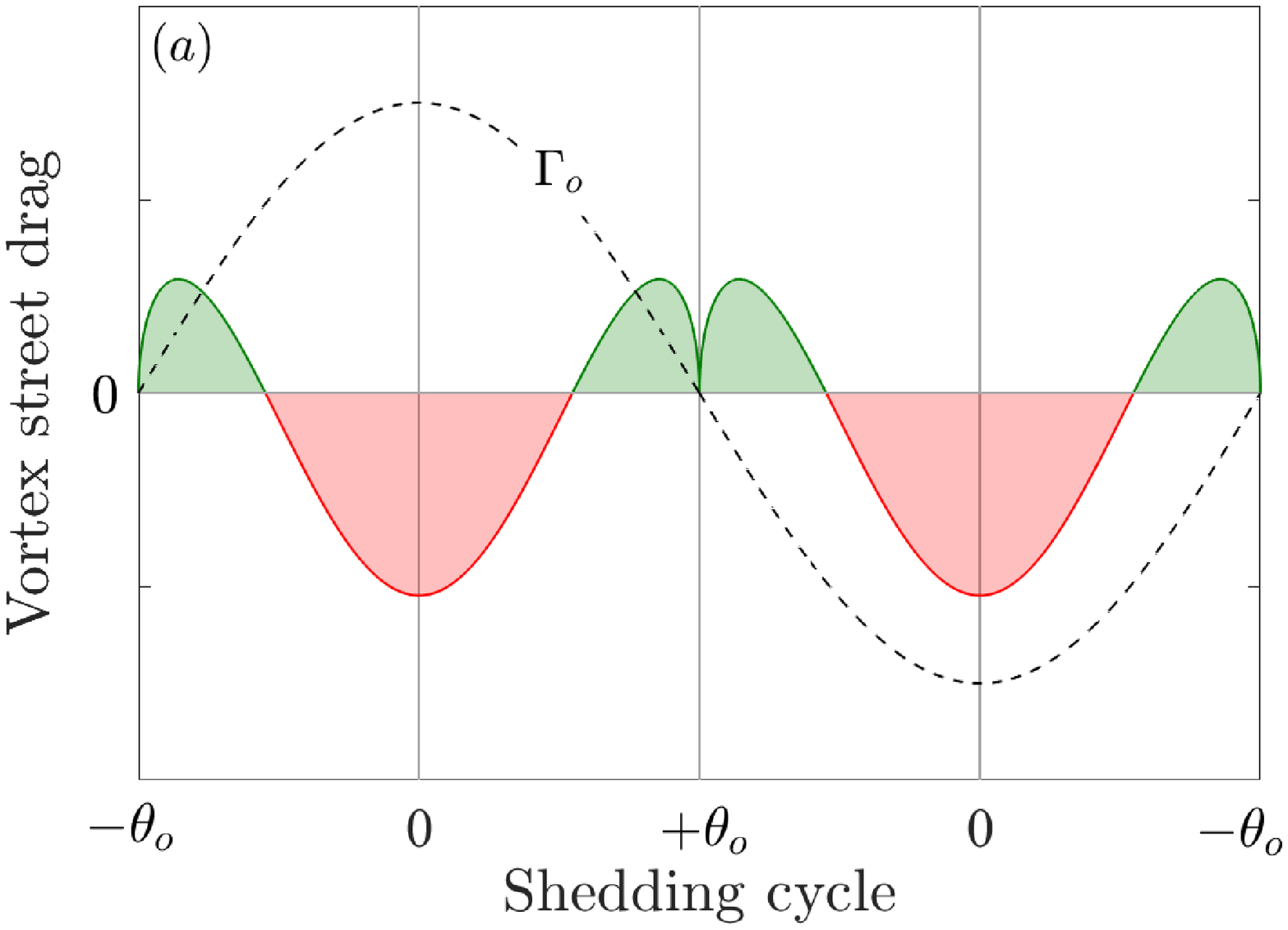}
\end{center}\end{minipage}
\begin{minipage}{0.49\linewidth}\begin{center}
\includegraphics[width=0.99\textwidth, angle=0]{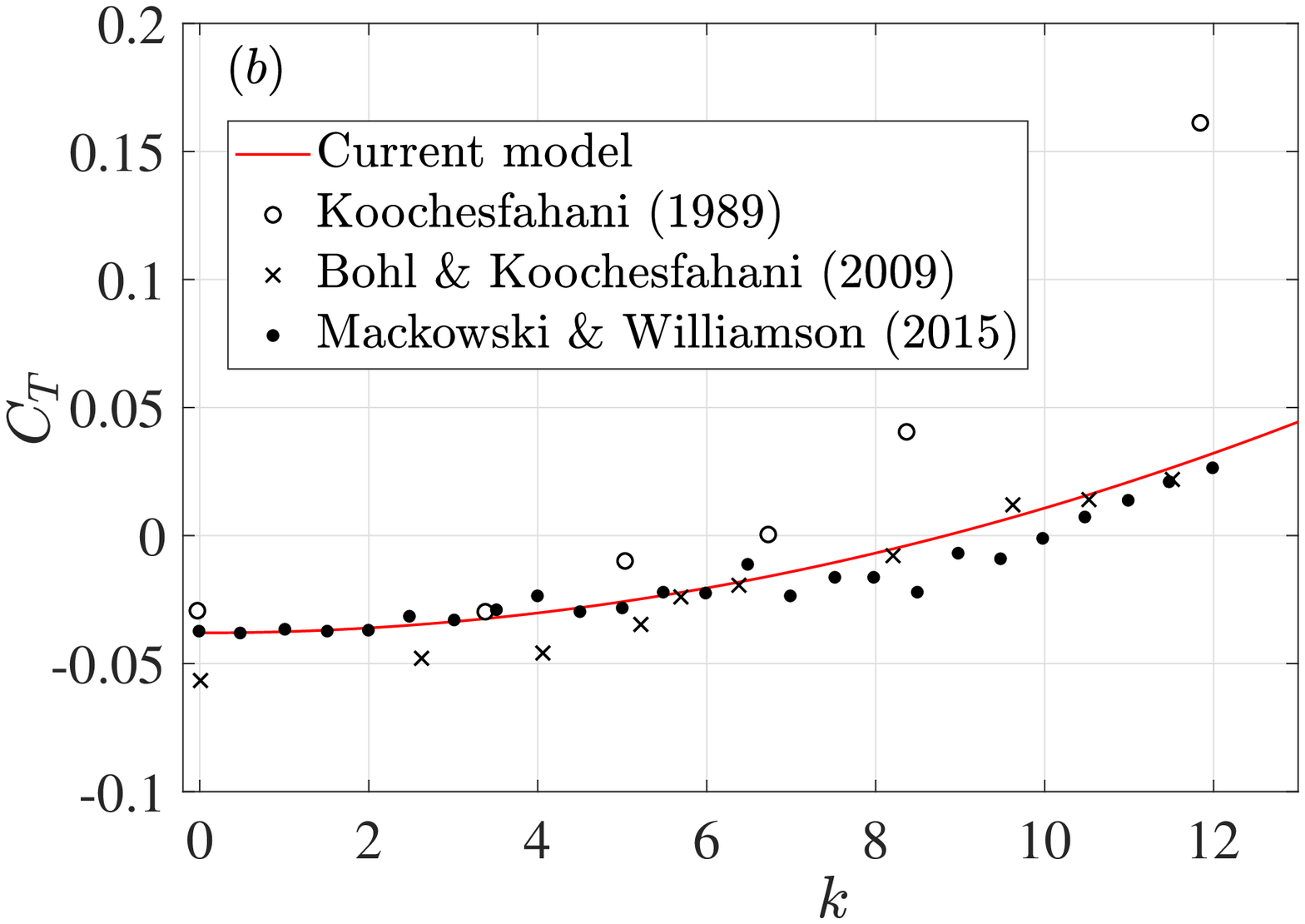}
\end{center}\end{minipage}
\caption{Modeled results for oscillating foils. (\textit{a}) Force history for vanishing amplitude $\theta_o\ll 1$ representing a von K{\'a}rm{\'a}n vortex street with a net drag. The circulation production (dashed black line) is in-phase with the acceleration and corresponds to the natural vortex shedding frequency. (\textit{b}) Modeled average thrust variation with $k$ (red line) for an oscillation amplitude $\theta_o=2^o$. Also shown are experimental measurements (symbols) with the same $\theta_o$.}
\label{fig:foil}
\end{center}
\end{figure}

From (\ref{eqn:Tmod}) we have $C_T=\bp{k\theta_oI_s}\Gamma_v^*$ where the non-dimensional characteristic vortex strength is $\Gamma_v^*=\Gamma_v/Uc$. Other theories predict a thrust coefficient with a frequency-squared dependency. Experiments confirm this behavior for oscillating foils with low amplitudes \citep[e.g.][]{Koochesfahani:89a,Anderson:98a,Koochesfahani:09a,WilliamsonCHK:15a}, and we infer that $\Gamma_v^*\propto k\theta_o$. Indeed, the linear trend with $k$ has been experimentally validated by the data of \cite{Koochesfahani:09a}. Therefore we write:
\begin{equation}\label{eqn:CT}
C_T=C_{T,0}+(k\theta_o)^2I_s\hspace{2pt}\der{\Gamma_v^*}{k}
\end{equation}
where $C_{T,0}<0$ is the static skin friction drag at $k=0$, and $\md{\Gamma^*}/\md{k}$ is the proportionality constant to be given from experiment. As $k$ increases, this drag is slowly overcome by the thrust generation. In the event that $f$ and $c$ are kept constant, this can be explained as follows. Here, the reduced frequency $k=2\pi f c/(2U)$ is decreased by increasing the flow speed $U$ in the facility. As such, this flow dominates the separation over the foil as it makes any excursion from the center line. Specifically, with a left-to-right free-stream, then foil positions above/below the center line generate shed vorticity of negative/positive sign. In the limit of zero amplitude, there are vortices of alternating sign very close to the center line. This is the configuration of a low Reynolds number von K{\'a}rm{\'a}n vortex street, namely a drag-inducing wake. Within the context of the current model, this can be represented by a circulation production with phase shift $\beta=\pi$, namely $\Gamma_o=\Gamma_v\cos\bp{\pi\theta_c/2\theta_o}$ so that $\pm\Gamma_v$ occur alternately on the center line. An example of the force history for this drag configuration is shown in figure~\ref{fig:foil}(\textit{a}) for a very small amplitude $\theta_o\ll 1$ and clearly exhibits a net drag. 

We can actually use this vortex street configuration to predict the skin friction drag value $C_{T,0}$. The Strouhal number is $St=fA/U$ where $f$ and $U$ are as before and $A$ is the peak-to-peak amplitude, which for small oscillations we can write as $A=2c\theta_o$. As such we have $k\theta_o=\frac{\pi}{2}St$. In the limit $\theta_o\rightarrow 0$ we take the Strouhal number to be that of the natural vortex shedding frequency, which for low Reynolds numbers is $St\approx 0.21$. This implies that $f$ becomes very large to keep $k\theta_o$ finite and we further assume that $k\theta_o=\Gamma_v^*$. Then computing the integral in (\ref{eqn:Tmod}) as $\theta_o\rightarrow 0$ with $\beta=\pi$ we find that $I_s\approx-0.35$ and finally $C_{T,0}\approx \bp{\frac{\pi}{2}St}^2I_s=-0.038$. This is remarkably close to experimental measurements which give $-0.6<C_{T,0}<-0.3$; see figure~\ref{fig:foil}.

As mentioned above, for $k>0$ the skin friction drag is overcome as the oscillation perturbs the wake configuration into a propulsive one. We calculate $C_T$ from (\ref{eqn:CT}) for the case of $\theta_o=2^o$ using $\md{\Gamma_v^*}/\md{k}\approx 0.04$, which was estimated from the data of \cite{Koochesfahani:09a}. The required integral in (\ref{eqn:Tmod}) was numerically computed with a circulation phase shift $\beta$ satisfying the zero-net side force condition on the half-cycle in (\ref{eqn:beta}). The result is plotted in figure~\ref{fig:foil}(\textit{b}) along with some experimental data and there is very favorable agreement. The phase shift $\beta$ was found to decay rapidly from $\pi/2$ to zero with increasing $k$. We recomputed $C_T$ using $\beta\equiv 0$ and the result was practically indistinguishable from that in figure~\ref{fig:foil}(\textit{b}).


For larger amplitudes more applicable to fish locomotion the relation (\ref{eqn:CT}) is not likely to hold. Ideally we still want an analytical expression for the thrust variation with the parameters of the problem. With our current choice of $\Gamma_o(\theta_c)$ as a trigonometric function we cannot evaluate the integrals in (\ref{eqn:Tfast}) and (\ref{eqn:Tmod}). However, if we allow `kinks' in $\Gamma_o$ by replacing the smooth trig functions with the following linear approximations then we can progress a bit further:
\begin{eqnarray}
\text{Asymmetric flapping:} && \Gamma_o=\Gamma_v\lrb{\frac{\theta_c}{\theta_o}} \label{eqn:Gasym} \\
\text{Symmetric flapping:} && \Gamma_o=\mp\Gamma_v\lrb{\frac{\bp{\theta_c\pm\theta_o}}{\theta_o}-\frac{2\beta}{\pi}}.
\end{eqnarray}
For the symmetric case, the upper sign is taken for $-\theta_o\leq\theta_c\leq 2\beta/\pi$ and the lower sign for $2\beta/\pi\leq\theta_c\leq \theta_o$. These are `saw-tooth' functions which, given the muscular agility of real fish, are arguably more accurate than the smooth trig functions. For the asymmetric case, the average thrust evaluates to 
\begin{equation}\label{eqn:Tfast2}
\frac{\overline{T}_\text{asym}}{\rho U\Gamma_v}=\frac{k\pi}{2}\lrb{H_0\bp{\theta_o}-\frac{H_1\bp{\theta_o}}{\theta_o}}\lrb{1-\frac{\Omega_\text{adv}}{\Omega_\text{ret}}}
\end{equation}
where $H_0(x)$ and $H_1(x)$ are the zero- and first-order Struve functions (non-homogeneous Bessel equation). The thrust is maximum at $\theta_o\approx 100^o$, which is very close to the value of $\theta_o=105^o$ reported in the live-fish experiments of \cite{EppsB:07a} for the angular displacement of the caudal fin of a Great Danio (\textit{Danio aequipinnatus}) performing a fast `C-start' turn in 0.25~s.

For the symmetric case the general integrals still cannot be evaluated analytically. A notable exception, however, is $\beta=0$, i.e. circulation production exactly out of phase with the velocity, for which we obtain:
\begin{equation}\label{eqn:Tsym}
\frac{\overline{T}_\text{sym}}{\rho U\Gamma_v} = \frac{k\pi}{4}J_1\bp{\theta_o} 
\end{equation}
where $J_1(x)$ is the Bessel function as before. If the reduced frequencies are equal, the symmetric flapping thrust in (\ref{eqn:Tsym}) is comparable with the asymmetric flapping thrust in (\ref{eqn:Tfast2}), depending on the stroke ratio $\Omega_\text{adv}/\Omega_\text{ret}$. However, the former is contingent on the condition $\beta=0$, which physically implies that there is no induced effect on the newly forming vortex at the start of the return stroke from the previously shed vortex of opposite sign. In non-emergency cruise conditions, a fish could approach this situation by slower undulation thus allowing the previously shed vortex to convect farther downstream. During predatory and/or escape maneuvers, however, the thrust production in (\ref{eqn:Tfast2}) by aggressive accelerations is more likely to be employed.

\section{Concluding remarks}\label{sec:conclude}
In this paper the calculation of the force on a body moving in an inviscid fluid was revisited with a fundamental difference concerning the vorticity generation due to the acceleration of the body. It was argued that, within the context of an inviscid model, acceleration of the body generates vorticity whose strength is independent of that required to satisfy the normal boundary condition. The remaining degree of freedom in the problem corresponds to the net effect of viscous stress on the body due to the enforcement of a tangential boundary condition in the higher-order Navier-Stokes equations governing the viscous fluid. More plainly, this degree of freedom is the net circulation established by the acceleration or, equivalently, around the body and any shed vorticity. Kelvin's circulation theorem requires that this circulation is immediately communicated to infinity. This is in contrast to the seemingly universal interpretation of Kelvin's theorem in applied problems as requiring this circulation to always be zero. We have shown that this assumption renders an inviscid model as incapable of capturing the viscous effect of vorticity generation due to acceleration of the body.

The calculation of the force experienced by a moving body was detailed in the presence of net circulation and mass entrainment on the boundaries, both of which complicate the usual process of employing the method of residues to obtain the Blasius integral. The entrainment represents the mass contained in viscous layers adjacent to the body or contained in the wake structure. The result also applies to surfaces with different mass density than the surrounding fluid. 

The derived force equation was demonstrated on several practical examples representative of biological propulsive modes that employ oscillatory mechanisms. In these cases, significant accelerations of the body/boundary occur and it was assumed that the associated vorticity generation dominated the force production, with induced wake effects considered secondary. It was shown that a non-zero average propulsive force requires some form of asymmetry. For fast and aggressive accelerations typical to predatory and escape maneuvers of fish, it was suggested that the caudal fin flaps asymmetrically in order to produce a net-positive thrust. This is because, under these circumstances, the circulation production is exactly out of phase with the acceleration and the fish actively utilizes the `wake capture' effect by pushing off of the wake momentum. For smooth, symmetric undulation typical to cruising, the asymmetry occurs as a phase shift in the circulation production relative to the fin motion. Here, wake capture is used actively and passively. The maximum thrust occurs when the circulation is exactly out of phase with the fin velocity. Physically, this corresponds to negligible effects from the previously shed vortex. The propulsive model was also compared to experiments on foils oscillating at low amplitude and showed good agreement with measured thrust values. 

\section*{Acknowledgments}
We wish to acknowledge the partial support of the NSF and ONR in this work.

\section*{Declaration of Interests}
The authors report no conflict of interest.

\bibliographystyle{jfm}
\bibliography{Inviscid_force_calc.bbl}
\end{document}